\documentclass[num-refs]{wiley-article}

\usepackage{siunitx}
\usepackage{hyperref}
\papertype{Original Article}
\paperfield{Journal Section}

\title{Predictability of north Pacific blocking events : Analogue based analysis of historical MIROC6 simulations}

\author[1,2]{Anupama K Xavier}

\author[1,2]{Oisín Hamilton}

\author[3,4,5]{Davide Faranda} 

\author[1]{Stéphane Vannitsem}

\affil[1]{Dynamical Meteorology and Climatology Unit, Royal Meteorological Institute of Belgium, Brussels, Belgium}

\affil[2]{Earth and Life Institute, Université catholique de Louvain, Louvain-la-Neuve, Belgium}

\affil[3]{Laboratoire des Sciences du Climat et de l’Environnement, Gif-sur-Yvette, Paris, France}

\affil[4]{London Mathematical Laboratory, London, UK}

\affil[5]{Laboratoire de Météorologie Dynamique/IPSL, Paris, France}
\corraddress{Anupama K Xavier, Dynamical Meteorology and Climatology Unit, Royal Meterological Institute of Belgium, Uccle, 1180 Brussels, Belgium}

\corremail{xavier.anupamak@meteo.be}

\fundinginfo{Horizon 2020 Framework Programme, Grant/Award Number: 956396}

\runningauthor{A. K. Xavier et al.}
\begin{document}
\maketitle
    
\begin{abstract}
    Atmospheric blocking exerts a profound influence on mid-latitude circulation, yet its predictability remains elusive due to intrinsic non-linearities and sensitivity to initial-conditions. While blocking dynamics have been extensively studied, the impact of geographical positioning on predictability remains largely unexplored. This study provides a comparative assessment of the predictability of Western and Eastern North Pacific blocking events, leveraging analogue-based diagnostics applied to CMIP6 MIROC6 simulations. Blocking structures are identified using geopotential height gradient reversal, with their temporal evolution analysed through trajectory tracking and error growth metrics. Results reveal that Eastern blocks exhibit lower predictability, characterized by rapid error divergence and heightened mean logarithmic growth rates, whereas Western blocks display dynamical stability. Persistence analysis gives no significant difference between eastern and western North Pacific blocking events. Sensitivity analyses across varying detection thresholds validate the robustness of these findings. 
\end{abstract}
    \section{Introduction}\label{sec:introduction}

Atmospheric blocking is a fundamental dynamical phenomenon that influences mid-latitude weather evolution by disrupting the typical westerly flow. Characterized by persistent, quasi-stationary high-pressure systems, these blocks can last from several days to weeks, leading to prolonged extreme weather conditions such as heatwaves, cold spells, and heavy precipitation events~\cite{rex1950blocking,tibaldi1990operational,masato2012wave,liu1994definition,nakamura2018atmospheric}. The North Pacific region is particularly prone to blocking events that play a key role in modulating weather and climate variability across North America and Eurasia~\cite{park2024hybrid,woollings2018blocking}.  

Despite its recognized impact on regional climate and extreme weather, the predictability of atmospheric blocking remains a challenge in numerical weather prediction (NWP) and climate modeling. The inherent nonlinear nature of blocking dynamics, along with their sensitivity to initial conditions and complex interactions with large-scale atmospheric waves, contributes to forecast uncertainty~\cite{masato2012wave,matsueda2018estimates,pelly2003new}. In the North Pacific, blocking events are typically categorized into two primary regions: Western North Pacific (WNP) and Eastern North Pacific (ENP) blocks~\cite{davini2012bidimensional}. Given the substantial downstream effects on weather and climate variability, enhancing the understanding of the predictability of these blocking regimes is crucial for improving forecast skill and weather model performance~\cite{drouard2019disentangling,merryfield2020current,breeden2020optimal}.  

Several studies have examined the predictability of North Pacific blocking using ensemble forecasting methods and model intercomparisons~\cite{matsueda2009blocking,schiemann2020northern,leutbecher2008ensemble}. While advancements in global climate models (GCMs) and numerical weather prediction systems have led to some improvements in simulating blocking characteristics, significant biases persist in representing their frequency, persistence, and decay~\cite{palmer2008toward,davini2016northern}. The latest generation of Coupled Model Intercomparison Project Phase 6 (CMIP6) models still struggle to capture high-impact extreme events linked to blocking, highlighting a critical gap in understanding the mechanisms governing their predictability~\cite{davini2020cmip3}.  

An overlooked aspect in existing research is the potential influence of the geographical position of a blocking event on its predictability. To date, no systematic studies have been conducted to determine whether the location of blocking relative to major topographical features affects its predictability. Research by~\cite{xavier2024variability}, based on a reduced-order atmosphere-land model~\cite{demaeyer2020qgs}, suggest that blocking events positioned to the west of a given topographical feature tend to exhibit lower predictability and greater instability compared to those occurring on the eastern side.

Building upon this work, the present study aims to investigate whether the predictability differences identified in idealized atmosphere-land coupled models are also evident in real-world simulations. Using CMIP6 MIROC6 model outputs, we analyze the predictability characteristics of North Pacific blocking events, particularly focusing on how their location relative to continental landmasses influences its predictability and stability. The spatial structure and geographical distribution of blocking events identified in~\cite{xavier2024variability} closely resemble North Pacific blocks, where high-pressure systems can occur either on the western or eastern side of major topographical features. In the physical world, these positions correspond to the Asian and American continents on either side of the Pacific, providing a basis for direct comparison with observational datasets.

The structure of the paper is as follows. Section~\ref{sec:data and methodology} introduces the data and methodology used for the investigation. In Section~\ref{sec:results}, properties of the reference-analogue pair is discussed along with the predictability properties of identified blocking events. The conclusions drawn from the research are provided in Section~\ref{sec:conclusions}, along with future perspectives for further studies.

    \section{Data and Methodology}\label{sec:data and methodology}
    In this section we outline the dataset and methodological framework that was used to analyse the predictability of atmospheric blocking in the North Pacific.

\subsection{Data}\label{subsec:Data}
    This study uses the output of the Coupled Model Intercomparison Project Phase 6 - Model for Interdisciplinary Research on Climate, version 6 (CMIP6 MIROC6) experiment~\citep{tatebe2018miroc}, in which the atmospheric component has a spatial resolution of approximately $1.4^\circ\times1.4^\circ$ (T85 spectral resolution, ~250 km), covering the period between 1850–2014, using the historical scenario. We selected the MIROC6 experiment output for this study due to its demonstrated capability in reproducing the climatological frequency of atmospheric blocking~\citep{palmer2023performance}.  MIROC6 has been shown to  exhibit smaller biases over the North Pacific~\citep{dolores2024role,davini2016northern}, made it suitable for assessing blocking dynamics in this region.  The analysis is based on $500\si{\hecto\pascal}$ geopotential height with a temporal resolution of one day. 

\subsection{Methodology}\label{subsec:methodology}

\begin{figure}[h]
        \centering
        \includegraphics[width=1\textwidth]{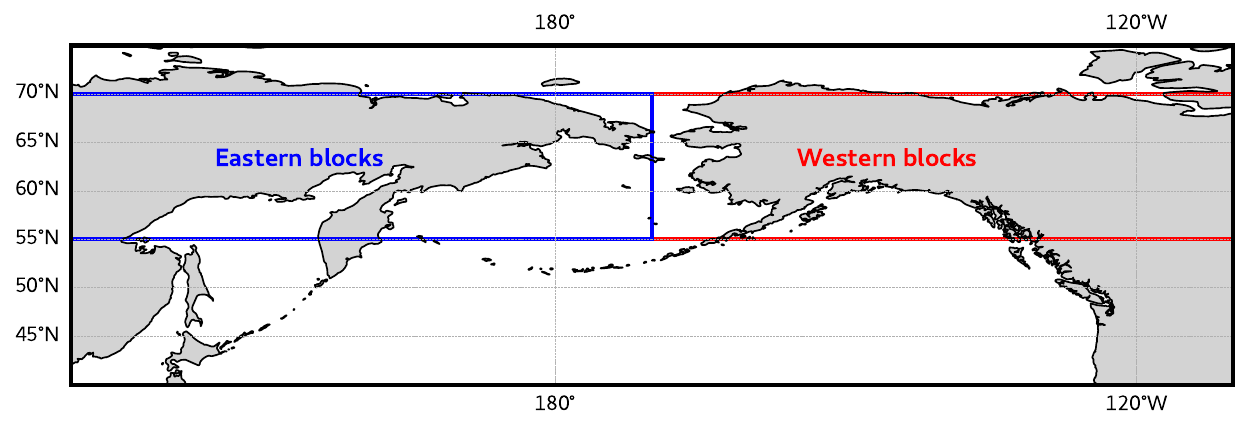} 
        \caption{Geographical domains used to define the Eastern and Western North Pacific blocking regions in this study. Both domains span latitudes of $55^\circ$N–$70^\circ$N. The Western North Pacific (WNP) region (blue) is bounded between longitudes of $190^\circ$E - $250^\circ$E, and the Eastern North Pacific (ENP) region (red) between $130^\circ$E–$190^\circ$E. These regions are highlighted by boxes to facilitate the identification of atmospheric blocking events.}
        \label{fig:pacific_regions}
    \end{figure}

    The initial step is to identify blocking events in the North Pacific region and classify these into eastern or western blocks based on the geographical location. The east and west domains were defined as boxes which is shown in Figure~\ref{fig:pacific_regions}.

\subsubsection{Identifying Blocking Events}\label{Identifying blocking events}

    To identify atmospheric blocking events, we employ the meridional gradient reversal method using the $500\si{\hecto\pascal}$ geopotential height (\( Z_{500} \)), following the approach introduced in~\citet{davini2012bidimensional}. This method identifies reversals in the geopotential gradient between a northern latitude (denoted GHGN) and a southern latitude (GHGS), and presented in Equation~\ref{eq:gph_gradients}.

    \begin{equation}\label{eq:gph_gradients}
        \begin{aligned}
            \text{GHGN}(t, \lambda_0, \Phi_0) &= \frac{Z_{500}(t, \lambda_0, \Phi_N) - Z_{500}(t, \lambda_0, \Phi_0)}{\Phi_N - \Phi_0}\\
            \text{GHGS}(t, \lambda_0, \Phi_0) &= \frac{Z_{500}(t, \lambda_0, \Phi_0) - Z_{500}(t, \lambda_0, \Phi_S)}{\Phi_0 - \Phi_S}
        \end{aligned}
    \end{equation}
    where \( \lambda_0 \) and \( \Phi_0 \) represent the longitude and latitude of the grid cell, respectively, and $t$ is the given time index. In this study we calculate blocking events at midday UTC on a daily basis, so the variable $t$ is an integer daily index.  \( \Phi_0 \) ranges from 30° to 75°N. \( \Phi_S \) = \( \Phi_0 \) - 15°, \( \Phi_N \) = \( \Phi_0 \) + 15°  and \( Z_{500}(t, \lambda_0, \Phi) \) denotes the geopotential height at $500\si{\hecto\pascal}$ at the specified latitude, on day $t$.

    A grid cell is considered to be experiencing instantaneous blocking (IB) if both of the following conditions hold:

    \begin{equation*}
        \begin{split}
            \text{GHGS}(t, \lambda_0, \Phi_0) &> 0
        \end{split}
        \quad
        \begin{split}
            \text{GHGN}(t, \lambda_0, \Phi_0)  &< -10 \si{\metre} (^\circ \text{lat})^{-1}
        \end{split}
    \end{equation*}

    \citet{davini2012bidimensional} introduced additional conditions to isolate blocking events that occur over large spatial scales, and have temporal persistence. This criterion ensures that only regions that exhibit a strong reversal in the meridional geopotential height gradient are considered blocking events, filtering out transient anomalies. To do this large scale blocks are identified by requiring that the IB condition is satisfied over a longitudinal span of at least $15^\circ$. 

    To include temporal persistence, a Blocking Event (BE) is detected if a cell that satisfies the large scale blocking condition occurs within a \( 5^\circ \) latitude \( \times \) \( 10^\circ \) longitude box centred on that grid cell, for at least five consecutive days. 
    
    Secondly, to avoid low latitude blocks in order to focus on mid latitude blocks and to avoid detecting block like features from tropics, this criteria is included 
    \begin{equation}
        \text{GHGS}_2(\lambda_0, \Phi_0) = 
        \frac{Z_{500}(\lambda_0, \Phi_S) - Z_{500}(\lambda_0, \Phi_S - 15^\circ)}{15^\circ}
    \end{equation}
    
    \begin{equation}
        < -5 \text{ m} (^\circ \text{lat})^{-1}.
    \end{equation}
    Hence a Blocking Event, $BE(t, x, y)$, is defined as follows

    \begin{equation*}
        BE(t, x, y) = \begin{cases}
            1 &\text{if meets criteria from \citet{davini2012bidimensional}}\\
            0 &\text{otherwise}\\
        \end{cases}
    \end{equation*}
    here $t$ is the time index and $x$ and $y$ is the longitude and latitude of the grid cell.
    
    Once the Blocking Events $(BE)$ are identified, the time-averaged blocking frequency at each grid cell over the entire analysis period is computed. This metric, denoted as \( AB(x,y) \), represents the proportion of time steps that a given grid cell is classified as blocked:

    \begin{equation*}
        AB(x, y) = \frac{\sum_{t\in T} BE(t, x, y)}{\# T},
    \end{equation*}

    where \( T \) is an index of all time steps, and $\#T$ represents the number of time steps considered.

    To quantify the regional mean blocking frequency \( BE_{\%} \), we compute the spatial average of \( AB(x, y) \) over the entire study domain, which is given by:

    \begin{equation*}
        BE_{\%} = \frac{\sum_{x,y} AB(x, y)}{\#x \cdot \#y}.
    \end{equation*}

    Here, $\#x$ and $\#y$ denote the total number of grid cells in the longitudinal and latitudinal directions, respectively. This metric provides an estimate of the overall blocking occurrence in the domain.

    A filtering process is then applied to retain only significant blocking occurrences. Grid cells where the local blocking frequency, \( AB(x, y) \) exceeds a fraction \( \alpha \) of the regional mean blocking frequency \( IB_{\%} \) are classified as significant blocking regions.

    \begin{equation*}
        F(x, y) =
        \begin{cases} 
            1, & \text{if } AB(x, y) > \alpha \cdot BE_{\%}, \\
            0, & \text{otherwise}.
        \end{cases}
    \end{equation*}

    The filtering coefficient \( \alpha \) is a dimensionless sensitivity parameter that allows for the comparison of blocking characteristics across different study areas. It ensures that the detection methodology is not biased by the size of the analysis domain. By applying this filtering criterion, we effectively select blocks that occur over the defined geographical area and have a spatial spread above a given threshold.
    
    In this study, a particular day is classified as a blocked day if the number of grid cells within the predefined study region satisfies the filtering condition based on a chosen threshold value of \( \alpha \). A given day \( t \) is considered a blocked day if the number of grid cells meeting the blocking condition exceeds a prescribed threshold for the selected \( \alpha \):
    
    \begin{equation*}
        \sum_{x,y} F(x,y) > N_{\alpha},
    \end{equation*}
    
    where \( N_{\alpha} \) represents the threshold number of grid cells that must satisfy the blocking condition for the day to be classified as blocked. After evaluating a range of threshold values for blocking day detection, the parameter \( \alpha \) was varied from 0 to 1 to identify the most suitable values for robust detection. Based on this analysis, \( \alpha = 0.20 \) and \( \alpha = 0.25 \) were selected as optimal thresholds, as they provide a balanced trade-off between capturing a sufficient number of blocking days and preserving the synoptic-scale spatial coherence of the blocking patterns. To assess the sensitivity of the detection methodology to threshold selection, the analysis was conducted independently for both \( \alpha = 0.20 \) and \( \alpha = 0.25 \). This dual approach enables a comparative evaluation of how variations in \( \alpha \) influence the identification and characteristics of blocking events.

\subsubsection{Tracking Blocking Events}\label{Tracking of blocking events}
    In the previous section we described how we identify Blocking Events $BE(t, x, y)$, however blocking events are defined on a particular day $t$ and provide no information about how they evolve to the next day $t+1$. To allow us to track the same blocking event through time we use the tracking methodology introduced in~\citet{schwierz2004}.
    
    For each day in the study period the grid cells that experience a blocking event ($BE(t,x,y)=1$) are grouped into connected clusters of grid cells. Two blocking events ($BE(t_0,x_1,y_1)=1$ and $BE(t_0,x_2,y_2)=1$) are defined as connected if a path exists between the two events which passes through grid cells $(x_i,y_i)$, where for all transiting grid cells $BE(t_0,x_i,y_i)=1$, and each grid cell shares an edge. In other words, two blocking events are in the same blocking cluster if they are connected to one another with other non-zero blocking events. These clusters were identified using the python library SciPy~\citep{virtanen2020}, which identifies connected regions in pixilated images~\citep{steinfeld2020}.  
    
\begin{figure}
        \centering
        \includegraphics[width=1\linewidth]{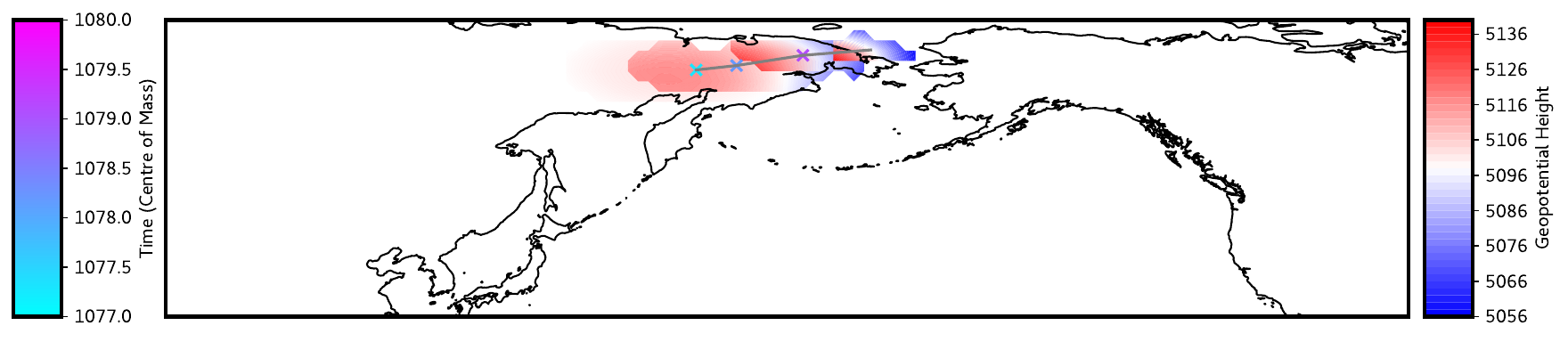}
        \caption{Schematic representation of tracking a blocking cluster identified on 1851-01-24 (index 388). `×' symbols indicate the centre of mass (CoM) of the cluster on each day, coloured by time progression (index) as shown by the left vertical colorbar. The background colour gradient represents geopotential height, with values indicated by the right vertical colorbar.}
        \label{fig:block_tracking_description}
    \end{figure}

    A blocking cluster at day $t+1$ is considered as the evolution of the cluster at day $t$ if they overlap sufficiently in space. This overlap is calculated by taking the spatial intersection of the clusters from one day to the next. The threshold is calculated by comparing the number of grid cells in the intersection with the number of grid cells in the cluster on the first day $t$. If the intersection covers at least half of the original cluster, we consider the cluster at $t+1$ to be the evolution of the cluster from time $t$. More concretely, let $C_i(t)$ be a cluster of blocking events at time $t$, and $C_j(t+1)$ another cluster on the next day. If $|C_i(t)\cap C_j(t+1)| / |C_i(t)|>0.5$ then we assume $C_i(t)$ evolves into $C_j(t+1)$. Here, we use $|\cdot|$ to denote the number of grid cells in each cluster. In this study we only track blocking clusters which have a single predecessor and a single successor cluster, thus removing any blocking clusters which split or merge from other blocks. An example of the process of tracking a particular blocking cluster is shown in Figure~\ref{fig:block_tracking_description}.

    Lastly, we calculate the centre of mass of each cluster, which provides a representative location for each block, enabling the identification of displacement and persistence while minimizing the influence of irregular block shapes and fragmented patterns. The centre of mass is weighted using the geopotential height. We apply this weighting as the centre of the blocking cluster as in this study we are using the geopotential height as the measure for identifying blocking events.

\subsubsection{Finding Analogues}\label{Finding analogues}

    Once blocked days are identified, we determine one analogue for each blocked day based on a similarity criterion. The analogue search is performed using a distance metric to compare atmospheric states.
    
    For the reference block, on day \( t_r \), its corresponding analogue day \( t_a \) is selected from the dataset based on a minimum distance criterion. The similarity between two atmospheric states is assessed using the spatial distribution of geopotential height at \(500 \, \si{\hecto\pascal}\). Here, \( X_t \) denotes the atmospheric state vector at time \( t \), represented by the gridded geopotential height field at \(500 \, \si{\hecto\pascal}\) over the selected spatial domain. It characterizes the synoptic-scale circulation pattern for a given day. The Euclidean norm is used to quantify the difference between the reference blocked day's geopotential height field and that of any candidate day as follows:
    
    \begin{equation}
        d(X_{t_r}, X_t) = \left\| Z_{500}(t_r) - Z_{500}(t) \right\|,
    \end{equation}
    
    where \( d(X_{t_r}, X_t) \) represents the Euclidean distance between the geopotential height fields at time \( t_r \) and \( t \). The dataset consists of spatial distribution of geopotential height values at \(500 \, \si{\hecto\pascal}\), and the goal is to identify \( t_a \) such that:
    
    \begin{equation}
        t_a = \underset{|t - t_r| \geq \Delta t_{\min}}{\operatorname{argmin}} \, d(X_{t_r}, X_t),
    \end{equation}
    
    where \( \Delta t_{\min} = 5 \), as in \citet{davini2012bidimensional}, which ensures a minimum temporal separation of at least 5 days between the blocked day and its analogue~\citep{diao2006new}. This constraint prevents the selection of analogues that are too close in time with the original reference block, thereby avoiding redundancy due to short-term atmospheric persistence.
    
    The analogue search procedure involves iterating over all available time steps, computing the Euclidean distance between the geopotential height distributions, and selecting the time step \( t_a \) that satisfies both the minimum distance criterion and the temporal separation condition. This approach ensures that each blocked day is assigned an analogue day that exhibits the most similar large-scale geopotential height distribution, providing insights into the recurrence of similar atmospheric circulation patterns.

    The traditional analogue method for assessing error evolution \cite{lorenz1969atmospheric} is not suitable in this context, as it fails to account for differences in predictability between eastern and western blocking regions due to the displacement of blocking Events following their initial identification. Therefore, error evolution based on the centre of mass is employed in the current study to better capture the spatial progression of blocking features.
    
    After identifying the reference and analogue blocking pairs, we track their temporal evolution using the block tracking algorithm, as explained in Section~\ref{Tracking of blocking events}. The trajectories of the reference and analogue blocks are denoted by \( \mathbf{X}_r(t) \) and \( \mathbf{X}_a(t) \), respectively, representing the latitude and longitude coordinates of the blocking centre over time.

\subsubsection{Measure of Predictability}

To assess the predictability of Eastern and Western North Pacific blocking events, we analyse the error evolution between the trajectories of the reference and analogue blocking pairs. The tracking of these blocking days is performed using a block tracking algorithm (Section~\ref{Tracking of blocking events}) that follows the movement of the centre of mass of each identified blocking system over time.

The error evolution between the reference and analogue blocking trajectories is computed using the Haversine distance, which measures the great-circle distance between two points on the Earth's surface:

\begin{equation}
    d_H(\lambda_r, \phi_r, \lambda_a, \phi_a) = 2 R \arcsin \left( \sqrt{\sin^2 \left( \frac{\phi_a - \phi_r}{2} \right) + \cos(\phi_r) \cos(\phi_a) \sin^2 \left( \frac{\lambda_a - \lambda_r}{2} \right)} \right),
\end{equation}

where \( d_H \) is the Haversine distance, \( R = 6371 \) km is the Earth's radius, and \( (\lambda, \phi) \) represent the longitude and latitude of the reference (\( \lambda_r, \phi_r \)) and analogue (\( \lambda_a, \phi_a \)) blocking centres.

For each pair of reference and analogue trajectories, we compute the error evolution at time \( t \) as:

\begin{equation}
    E_i(t) = d_H\left( \mathbf{X}_{r,i}(t), \mathbf{X}_{a,i}(t) \right),
\end{equation}

where \( E_i(t) \) represents the displacement error for the \( i \)-th reference–analogue pair at time step \( t \). The errors are recorded for all analogue pairs and stored in an error matrix, allowing for further statistical analysis.

To examine the variability of predictability, the initial errors at \( t_0 \) (the initial time) are sorted in ascending order, and the first 50 and 75 analogue pairs, with the smallest initial displacement errors, are selected for further analysis. These pairs are chosen because the initial separation between the reference and analogue trajectories is small, thereby ensuring a close initial state. This enhances the robustness of the error growth analysis by minimizing the influence of initial condition differences and allowing for a meaningful assessment of dynamical divergence over time.

The statistical distribution of errors is analysed by computing the mean and standard deviation of the error evolution across all reference-analogue pairs. To estimate the uncertainty in the error evolution, a bootstrap resampling technique is applied, and the 95\% confidence interval (CI) is obtained.

To assess the exponential growth of initial uncertainties, we compute the logarithmic growth rate for each pair at lead time \( \tau \) and increment \( \Delta \tau \), which quantifies the sensitivity of blocking events to small initial perturbations:

\begin{equation}
    \lambda_i(\tau) = \frac{1}{\Delta \tau} \ln \left( \frac{E_i(\tau + \Delta \tau)}{E_i(\tau)} \right),
\end{equation}

where \( E_i(\tau) \) is the displacement error for the \( i \)-th pair at lead time \( \tau \), and \( \Delta \tau \) is the time increment.

The mean logarithmic growth rate at each lead time is then computed by averaging over all analogue pairs:

\begin{equation}
    \bar{\lambda}(\tau) = \frac{1}{N} \sum_{i=1}^{N} \lambda_i(\tau),
\end{equation}

where \( N \) is the total number of analogue realizations. Comparing \( \bar{\lambda} \) for Eastern and Western North Pacific blocks reveals differences in their intrinsic predictability characteristics.

\subsubsection{Persistence of Blocking Events}
    To quantify persistence, we define the lifetime of a blocking event based on its tracked trajectory over time. Given a blocking system that is identified at time \( t_0 \) and remains detectable until time \( t_f \), its persistence \( P \) is given by $P = t_f - t_0 + 1$, where \( t_0 \) is the initial and \( t_f \) is the final time step where the block is detected.

    This methodology enables the analysis of the persistence of blocking events, providing ideas into their stability and potential for long-term atmospheric impacts.
    Overall, the analysis above provides both a deterministic and probabilistic measure of stability through the persistence and mean logarithmic growth rate computation allowing for a comprehensive assessment of the stability and comparability of north Pacific blocking events.
    \section{Results and Discussion}\label{sec:results}
    \subsection{Climatology of Eastern and Western North Pacific blocks}\label{Climatology of Eastern and Western North Pacific blocks}
\begin{figure}[h]
                \centering
                \includegraphics[width=1\textwidth]{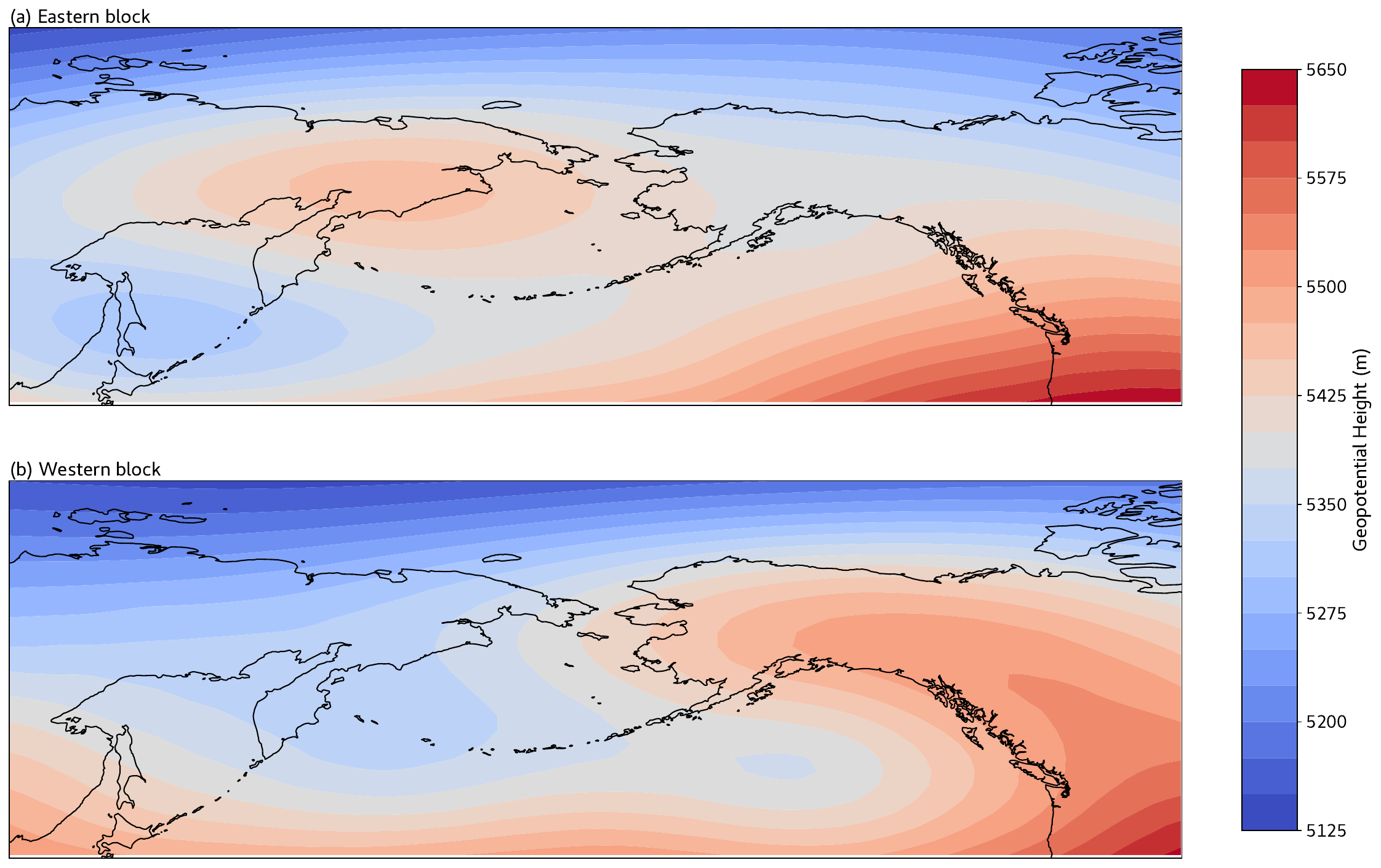} 
                \caption{Spatial distribution of the climatological geopotential height (m) for \textbf{(a)} Eastern blocks and \textbf{(b)} Western blocks, obtained using the methodology described in section~\ref{sec:data and methodology} with $\alpha = 0.25$. The climatology is computed over the base period from 1850 to 2014.}
                \label{fig:east_west_climatology_25}
            \end{figure}

        The climatological distribution of Eastern and Western blocks, computed using the methodology detailed in Section~\ref{subsec:methodology} with $\alpha = 0.25$, is presented in Figure~\ref{fig:east_west_climatology_25}. The classification effectively distinguishes the Eastern North Pacific (ENP) and Western North Pacific (WNP) blocking patterns. The spatial distribution of geopotential height reveals that Eastern blocks are primarily concentrated over the Aleutian region of Russia. This indicates that the most identified Eastern blocks exhibit blocking highs in this region. Conversely, Western blocks exhibit a more extensive distribution, with blocking highs dispersed across Alaska. The Western blocks are characterized by an intense high-pressure system, as evident from the deeper red-shaded regions, compared to the Eastern blocks. While the Eastern block presents a confined structure, the Western block's influence extends over a broader region of the North Pacific. This spatial distinction suggests potential differences in the dynamical characteristics and downstream impacts of these blocking patterns.

\begin{figure}[h]
                \centering
                \includegraphics[width=1\textwidth]{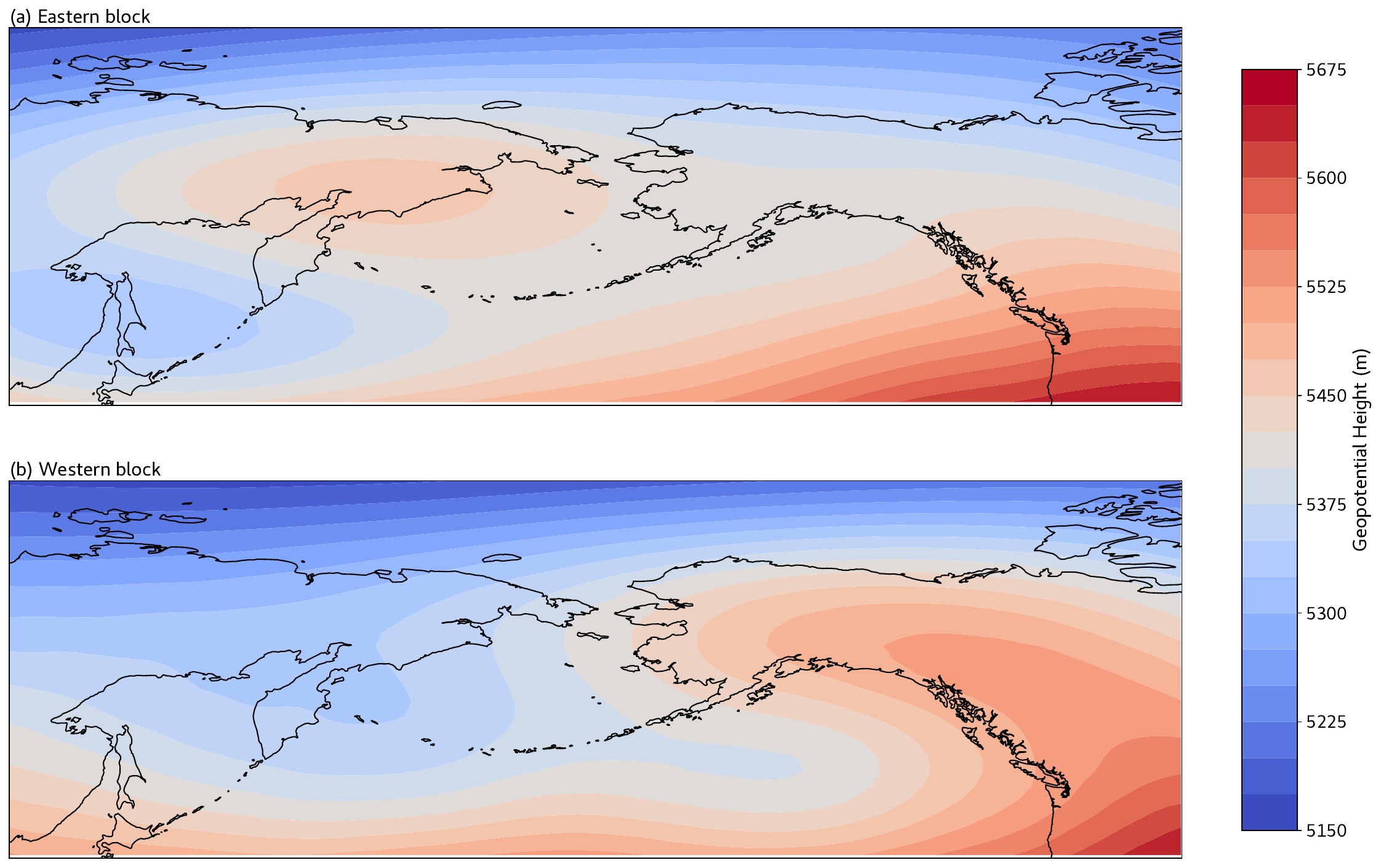} 
                \caption{Same as fig~\ref{fig:east_west_climatology_25} but with $\alpha = 0.20$.}
                \label{fig:east_west_climatology_20}
            \end{figure}

        In Figure~\ref{fig:east_west_climatology_20}, the climatological distribution of blocking events is shown for $\alpha = 0.20$. The spatial pattern of geopotential height anomalies closely resembles that observed in Figure~\ref{fig:east_west_climatology_25}. However, the number of identified blocking events is increased due to the decreased value of $\alpha$, which imposes a more lenient criterion for block detection.
 
    \subsection{Persistence of Eastern and Western blocks}

        Persistence of the Eastern and Western blocks were calculated for $\alpha = 0.20$ and $\alpha=0.25$. 

            \begin{figure}[h]
                \centering
                \includegraphics[width=1\textwidth]{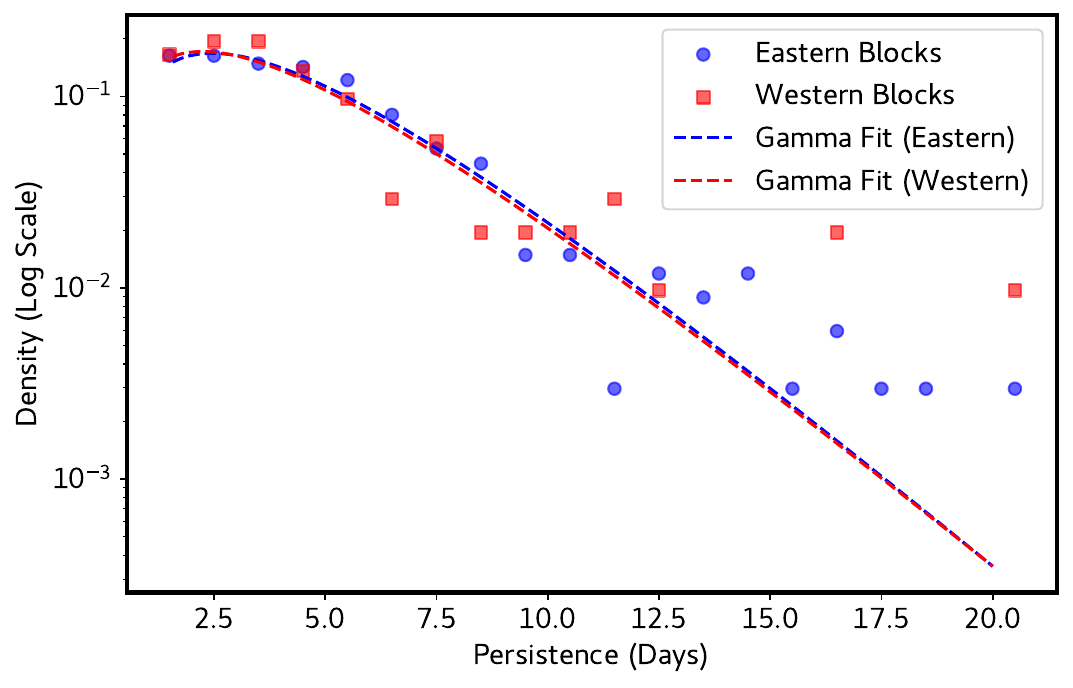} 
                \caption{Log-scaled density distribution of blocking persistence (in days) for Eastern and Western blocks with $\alpha = 0.25$. The fitted gamma distributions for both cases are also shown.}
                \label{fig:Persistance_25}
            \end{figure}
        Figure~\ref{fig:Persistance_25} presents the probability density distribution of blocking persistence, with a fitted gamma distribution applied to the entire dataset. The initial portion of the distribution exhibits a strong overlap between Eastern and Western blocks, particularly for block lifetimes up to 10 days. Beyond this threshold, the tails of the distributions diverge significantly, indicating a higher occurrence of long-lived Eastern blocks compared to Western blocks. Eastern blocks demonstrate a tendency to persist beyond 10 days, whereas Western blocks exhibit a sharper decline in frequency for extended durations.

            \begin{figure}[h]
                \centering
                \includegraphics[width=1\textwidth]{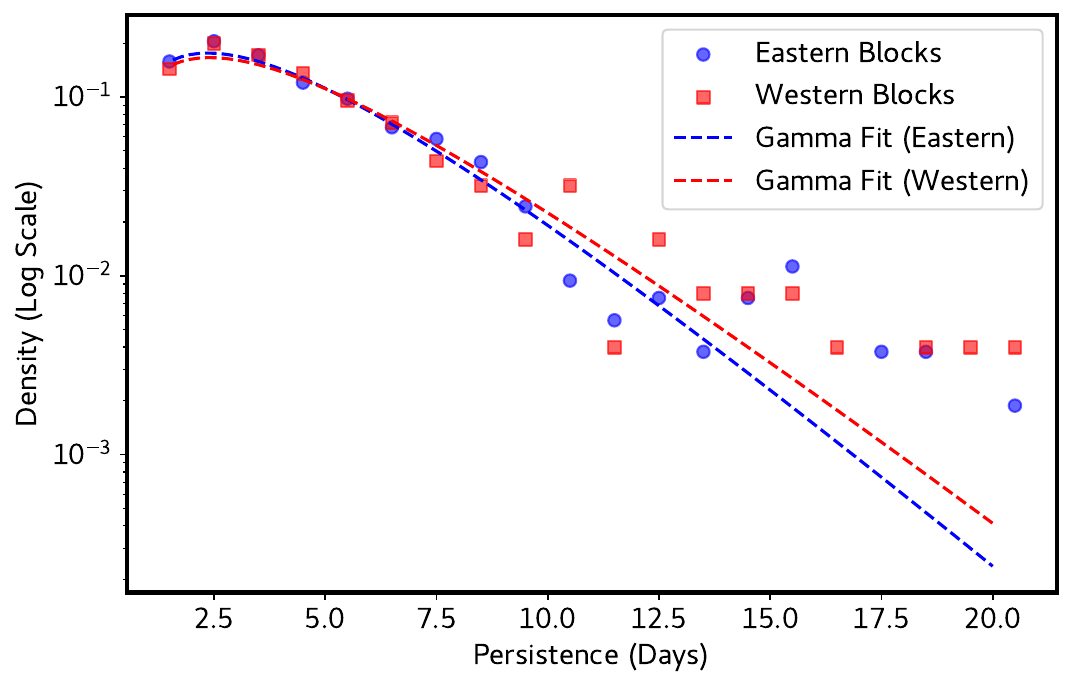} 
                \caption{Same as Figure~\ref{fig:Persistance_25} but with $\alpha=0.20$}
                \label{fig:Persistance_20}
            \end{figure}

        When the sample size is increased by reducing the selection threshold to $\alpha=0.20$ (Figure~\ref{fig:Persistance_20}), the initial portion of the distribution remains substantially unchanged. However, the disparity in long-lived blocking events diminishes, with Eastern and Western blocks displaying a more balanced distribution of extended persistence. 

        The observed changes with decreasing $\alpha$ can be attributed to the inclusion of a broader range of blocking events, incorporating shorter-lived and less persistent blocks that were previously excluded under a stricter selection criterion. Given the constraints of our methodology, we did not identify notable differences in the persistence between eastern and western North Pacific blocking events.
        
    \subsection{Structural Comparison of Reference and Analogue Blocks}\label{Structural Comparison of Reference and Analogue Blocks}
\begin{figure}[h]
                \centering
                \includegraphics[width=1\textwidth]{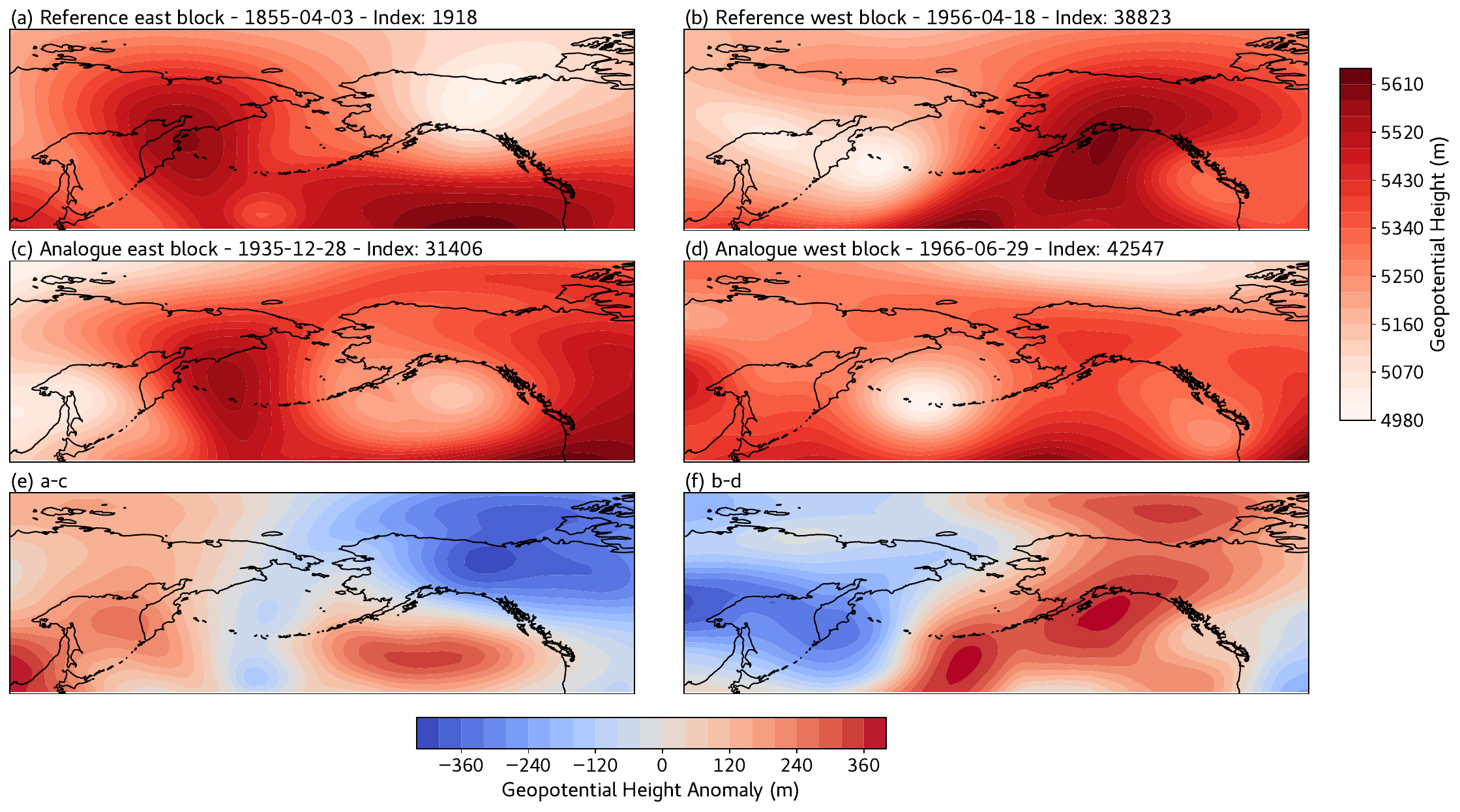} 
                \caption{\textbf{(a)} and \textbf{(b)} represent the geopotential height distribution of the reference Eastern and Western blocks, respectively, while \textbf{(c)} and \textbf{(d)} correspond to their analogue blocks. \textbf{(e)} and \textbf{(f)} depict the geopotential height anomalies between the typical and analogue blocks for the Eastern and Western cases, respectively.}
            \label{fig:Examples_east_west_blocks_anomaly}
            \end{figure}

        This section examines the geopotential height distribution of selected typical Eastern and Western blocks and their corresponding analogue blocks.
        In Figure~\ref{fig:Examples_east_west_blocks_anomaly} spatial distribution of geopotential height of a selected Eastern and Western block and its corresponding analogues were depicted. In order to identify the difference in the distribution, anomaly between them is also computed. 
            
        Figure~\ref{fig:Examples_east_west_blocks_anomaly}a and Figure~\ref{fig:Examples_east_west_blocks_anomaly}c represent the typical-analogue pair for the Eastern block, while Figure~\ref{fig:Examples_east_west_blocks_anomaly}b and Figure~\ref{fig:Examples_east_west_blocks_anomaly}d correspond to the Western block pair. Both reference-analogue pairs exhibit a high degree of similarity in their geopotential height distributions, indicating that the analogue blocks effectively capture the large-scale structure of the reference blocks. The differences between the selected reference and analogue blocks, depicted in Figures~\ref{fig:Examples_east_west_blocks_anomaly}e and \ref{fig:Examples_east_west_blocks_anomaly}f, reveal anomalies within a range of $\pm720\si{\meter}$, which is significantly smaller than the typical anomaly range of $\pm 1000\si{\meter}$ observed between two randomly selected blocking days. This suggests that the analogue blocks resemble their respective reference blocks much more closely, with only minor variations in intensity and spatial extent. The anomaly patterns highlight localized differences, which may be attributed to transient synoptic-scale features or differences in block persistence. These findings reinforce the reliability of the analogue approach in identifying predictability and stability of atmospheric blocking patterns.

    \subsection{Stability and predictability of identified blocks}\label{Stability and predictability of identified blocks}   
       The fundamental concept of assessing atmospheric predictability through analogue-based error diagnostics has been widely discussed in older meteorological literature \cite{lorenz1969atmospheric}. This study builds upon these foundational ideas to evaluate and compare the predictability of Eastern and Western North Pacific blocking events. A key metric in those analyses were the reference-analogue error, quantified using the weighted root-mean-square (RMS) height difference, which measures deviations between two atmospheric states.  Reference-analogue errors arise due to inherent dynamical differences between effectively occurring atmospheric situations, in particular for blocking events with similar large-scale structures but varying temporal evolution \cite{yiou2007inconsistency,faranda2017dynamical,faranda2023atmospheric}. In earlier studies, the progression of analogue errors has been recognized as a diagnostic tool for assessing the stability of blocking regimes, distinguishing between persistent and transient events \cite{plaut1994spells}. Furthermore, analogue-based diagnostics have  been utilized to study large-scale flow recurrence, offering an alternative to ensemble forecast approaches \cite{yiou2014anawege}. 

            \begin{figure}[h]
                \centering
                \includegraphics[width=1\textwidth]{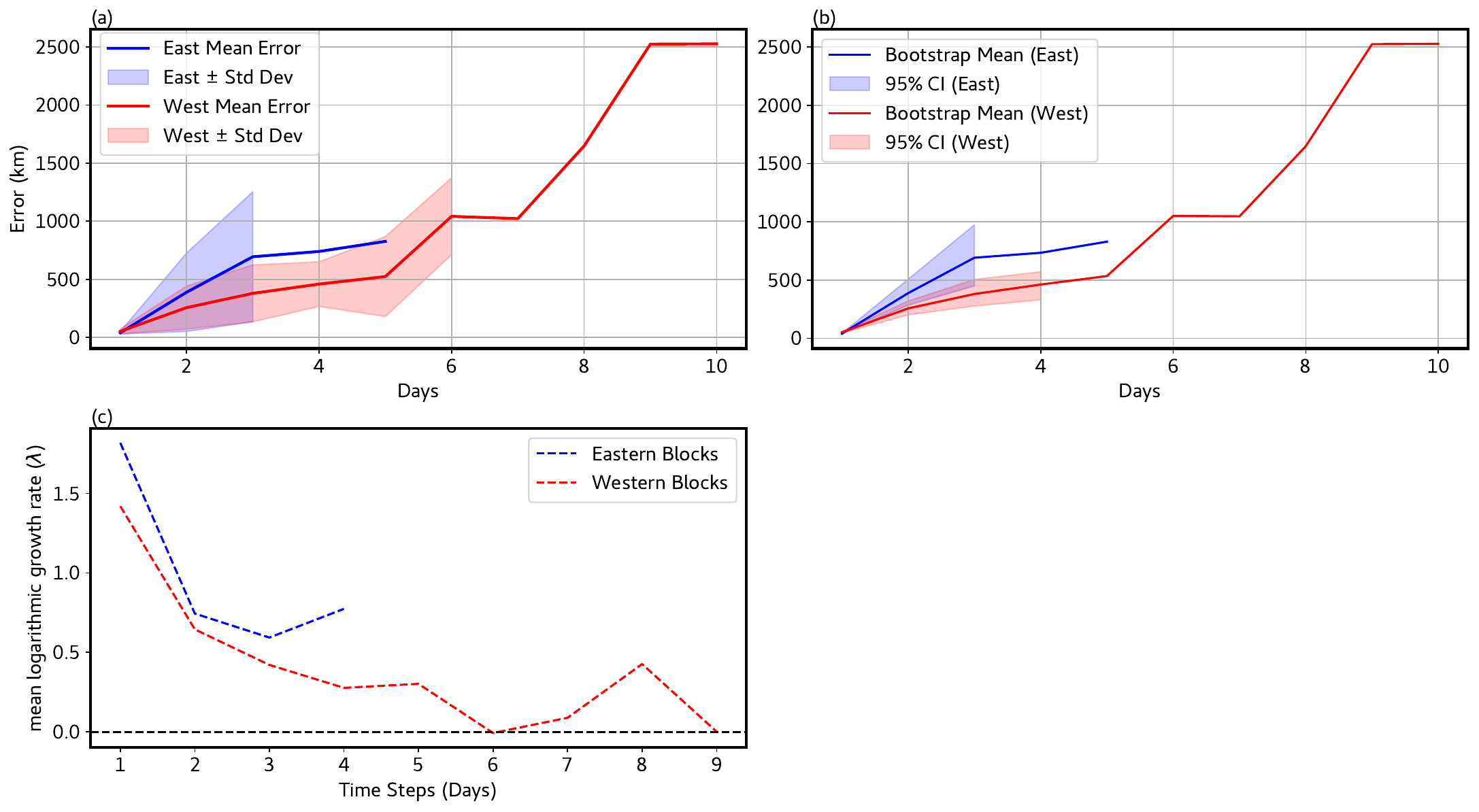} 
                \caption{Evolution of error and logarithmic growth rate for Eastern and Western blocks, considering $\alpha = 0.25$ and the first 50 reference-analogue pairs with the smallest initial error. \textbf{(a)} Mean error growth over time for Eastern and Western blocks, with shaded regions representing the standard deviation. \textbf{(b)} Bootstrap mean error with 95\% confidence intervals for both block types. \textbf{(c)} Lyapunov exponent evolution over time, comparing the stability characteristics of Eastern and Western blocks.}
                \label{fig:Error_lyapunov_25_50}
            \end{figure}
        According to \citep{lorenz1969atmospheric}, an analogue exhibits the same spatial distribution as the reference block plus a deviation. By selecting an analogue that minimizes the distance with their reference blocks, this initial deviation is the smallest, and the comparison of their subsequent evolutions provide an estimate of the local (in phase space) predictability of the reference block. To ensure a robust assessment, the error between trajectories is computed for all reference-analogue pairs in the dataset. From this large set, the 50 and 75 cases with the smallest initial deviation are selected, ensuring that the chosen blocks have minimal separation in the two-dimensional space, where the position of each block is determined by the centre of mass weighted by geopotential height.
       
       Figure~\ref{fig:Error_lyapunov_25_50}a illustrates the evolution of the average error for the first 50 reference-analogue pairs, quantifying the divergence between their trajectories. The shaded regions denote the standard deviation, capturing the variability in error growth across different blocking cases. The results indicate that the trajectories of Eastern blocks diverge  more rapidly than those of Western blocks, suggesting that Eastern blocks exhibit lower stability. Figure~\ref{fig:Error_lyapunov_25_50}b presents the bootstrap mean error for both Eastern and Western blocks, along with their respective 95\% confidence intervals (CIs). The use of the bootstrap method (with replacement) enhances statistical robustness by resampling the data to estimate uncertainty. The widening confidence intervals over time reflect an increasing separation between reference and analogue blocks, reinforcing the idea that blocking predictability diminishes with lead time. This decline is pronounced for Eastern blocks, indicating a greater rate of divergence compared to Western blocks. Figure~\ref{fig:Error_lyapunov_25_50}c depicts the averaged growth rate of the logarithm of the error, which is equivalent to the definition of the Lyapunov exponents, which characterizes how quickly small perturbations in initial conditions amplify, offering insight into the chaotic behaviour of atmospheric blocking. Higher values correspond to increased instability and reduced predictability. Note that in the case of a pure exponential behaviour along the dominant local instability, this averaged growth rate would have been constant as a function of time. 
       
       The results suggest that Eastern blocks exhibit larger logarithmic growth rate, implying a larger sensitivity to initial perturbations and, consequently, lower predictability compared to Western blocks. Over time, the growth rate decreases for both cases, that could indicate either the progressive saturation of the error as expected when the error becomes large \cite{nicolis1995short,vannitsem1997lyapunov}, or the increase of predictability of the blocking after its onset. Disentangling one from the other would need to make model experiments with very small initial condition errors at different stages of the development of blocking. The analysis above however indicates the predominant instability of Eastern blocks over Western blocks. 

        An important comment is in order here concerning the nature of the error growth in the position of the blocking events just illustrated here. As discussed in details in \citet{bohr1998dynamical}, one should distinguish between the Eulerian and Lagrangian Lyapunov instabilities. The first one is associated with the difference in the initial state of the background flow, while the second related with the spatial difference between the positions of the structures under interest, i.e the blocking events. Here both Lagrangian and Eulerian errors are affecting the dynamics, and one cannot distinguish between the two here as these are intertwined. In \citet{bohr1998dynamical}, it is also shown that in the presence of both errors, the largest exponent among the Lagrangian and Eulerian Lyapunov instabilities is controlling the error evolution. We thus suspect that in our case the maximum Lyapunov exponent between the Eulerian and Lagrangian approaches is at play. The first value of Figure~\ref{fig:Error_lyapunov_25_50}c can then be viewed as an estimate of the dominant local (in physical space) Lyapunov instability of the blocking events.

            \begin{figure}[h]
                \centering
                \includegraphics[width=1\textwidth]{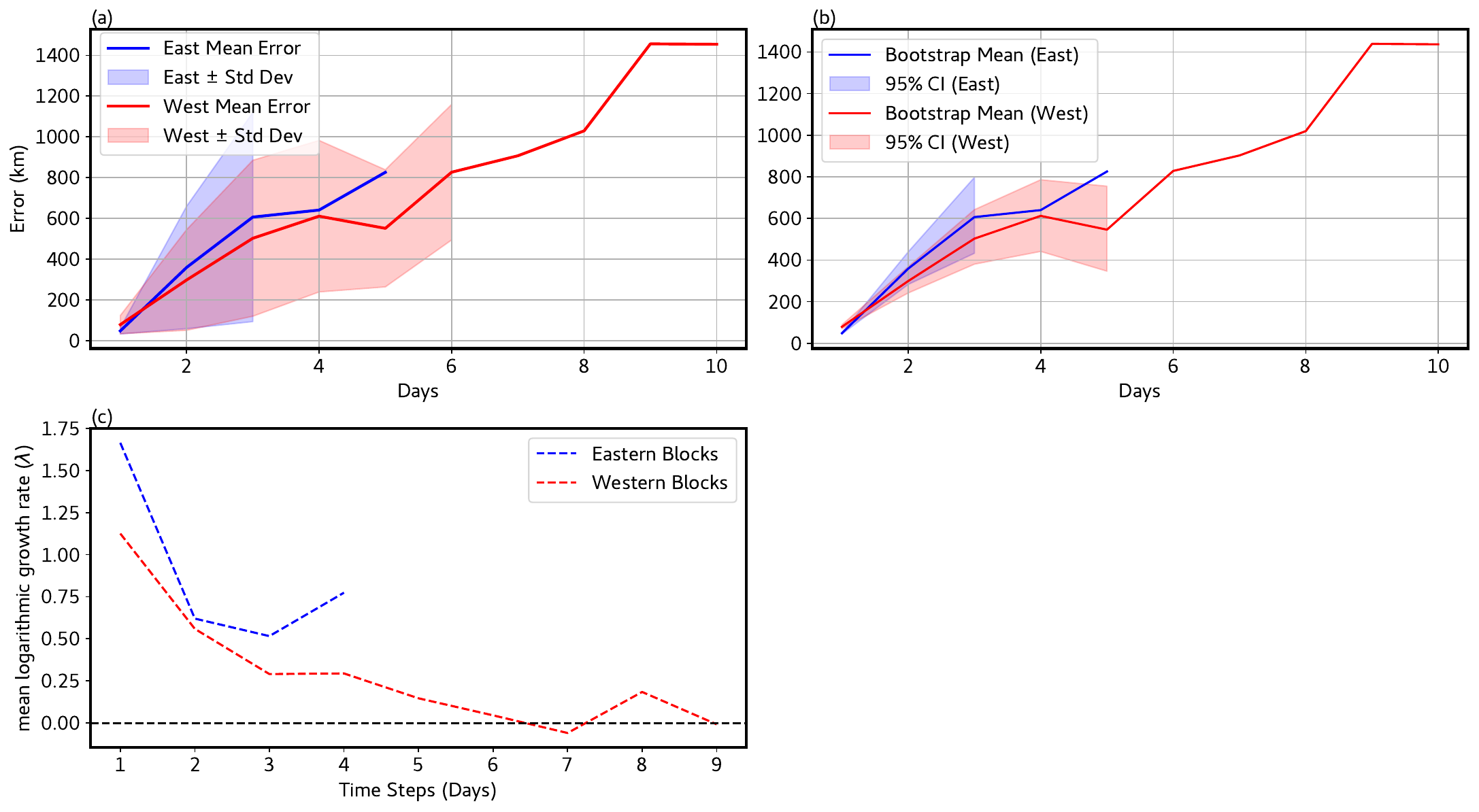} 
                \caption{Same as Figure~\ref{fig:Error_lyapunov_25_50} but with $\alpha=0.25$ considering first 75 reference-analogue pairs}
                \label{fig:Error_lyapunov_25_75}
            \end{figure}
        
        The same analysis was repeated for the first 75 reference–analogue pairs, as shown in Figure~\ref{fig:Error_lyapunov_25_75}. In Figures~\ref{fig:Error_lyapunov_25_75}a and b, the average error evolution for Eastern and Western blocks appears nearly indistinguishable, indicating reduced contrast compared to Figures~\ref{fig:Error_lyapunov_25_50}a and b. However, a slight difference remains, with Eastern blocks still exhibiting a marginally higher rate of error growth than their Western counterparts. The variability, quantified as the spread in error evolution across the reference–analogue pairs --calculated using both the standard deviation and the bootstrap 95\% confidence interval --overlaps for both cases. This reduction in the difference between the average error evolution suggests that increasing the sample size introduces greater diversity in blocking characteristics with larger initial condition errors, leading to a broader representation of stability and transition behaviours. As a result, the initially observed contrast between Eastern and Western blocks diminishes, highlighting the influence of sample selection on assessing blocking predictability. Mean logarithmic growth rate in Figure~\ref{fig:Error_lyapunov_25_75} shows comparable difference for both cases by highlighting lower predictability for Eastern blocks as same as in Figure~\ref{fig:Error_lyapunov_25_50}. Despite increasing the sample size from 50 to 75 cases, the logarithmic growth rate continue to exhibit significantly different values for Eastern and Western blocks. This suggests that the inherent instability of Eastern blocks is a fundamental characteristic, less dependant on the number of selected cases. 
            \begin{figure}[h]
                \centering
                \includegraphics[width=1\textwidth]{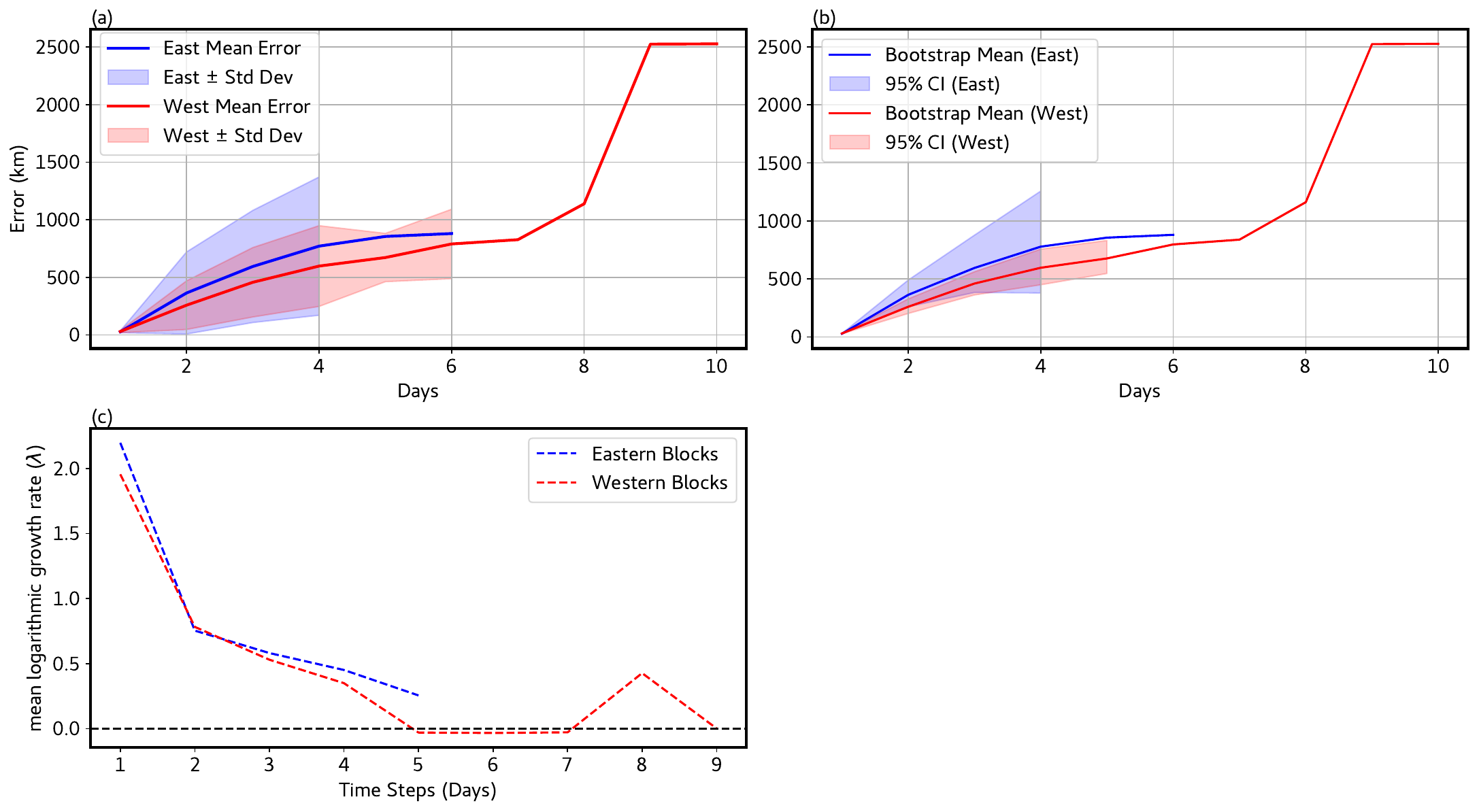} 
                \caption{Same as Figure~\ref{fig:Error_lyapunov_25_50} but with $\alpha=0.20$ considering first 50 reference-analogue pairs}
                \label{fig:Error_lyapunov_20_50}
            \end{figure}

        The analysis was repeated for $\alpha = 0.20$, and the resulting error evolution and mean logarithmic growth rate for the first 50 cases are presented in Figure~\ref{fig:Error_lyapunov_20_50}. The mean error evolution exhibits a similar pattern to previous analyses, with higher values for Eastern blocks, reaffirming their lower stability. However, the logarithmic error growth rate (Figure~\ref{fig:Error_lyapunov_20_50}c) shows a distinct behaviour: initial differences between Eastern and Western blocks gradually diminish, leading to an overlap on day 2. As time progresses, the divergence becomes significant again, particularly on days 3, 4, and 5, further supporting the conclusion that Eastern blocks are inherently more unstable than Western blocks.
   
            \begin{figure}[h]
                \centering
                \includegraphics[width=1\textwidth]{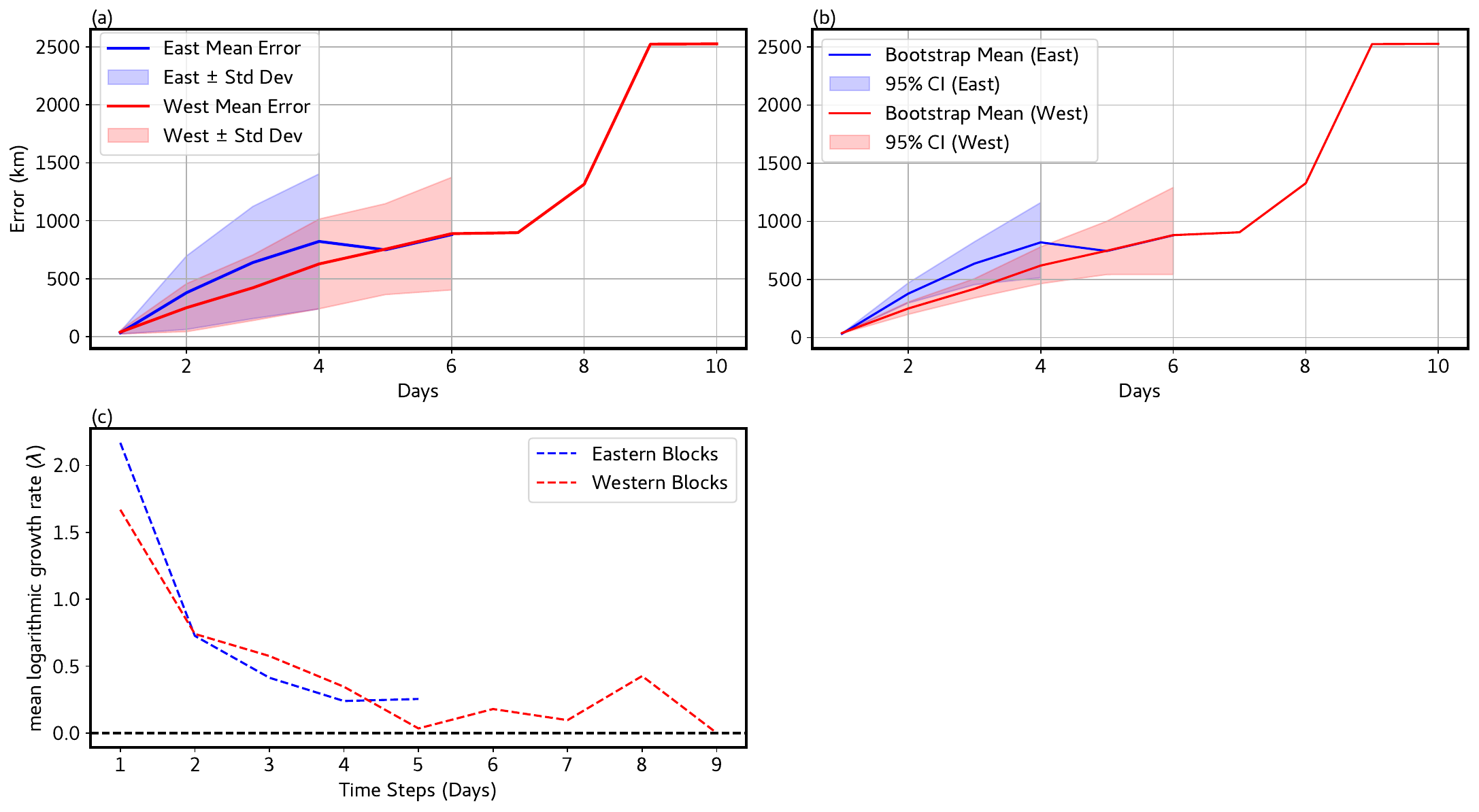} 
                \caption{Same as Figure~\ref{fig:Error_lyapunov_25_50} but with $\alpha=0.20$ considering first 75 reference-analogue pairs}
                \label{fig:Error_lyapunov_20_75}
            \end{figure}

        For the first 75 cases (Figure~\ref{fig:Error_lyapunov_20_75}), a similar pattern emerges. Initially, Eastern blocks exhibit higher error values, but beyond the fifth day, the error evolution of Eastern and Western blocks becomes more comparable, leading to an overlap. The logarithmic error growth rate follows a non-monotonic behaviour: higher initial values for Eastern blocks, followed by relatively lower values on days 3 and 4, and a subsequent increase on day 5.
        
        Reducing the threshold \( \alpha \) from 0.25 to 0.20 increases sensitivity to smaller-scale variability by allowing a broader range of blocking events. This broader inclusion introduces greater variability in error growth patterns, as reflected in the non-monotonic behaviour of logarithmic error on days 3 and 4. With a less stringent analogue selection, initial differences between reference and analogue blocks are slightly larger, which may explain overlapping growth phases—such as day 2 for the 50-case scenario and day 5 for the 75-case scenario—before divergence occurs. These effects are further amplified in larger samples, where a mix of more persistent and less stable blocks leads to alternating phases of stronger and weaker error growth.

        In contrast, using a stricter threshold of \( \alpha = 0.25 \) ensures the selection of only the most structurally similar blocking events, making the differences in their instability characteristics more apparent. Despite the added variability introduced at \( \alpha = 0.20 \), the underlying instability of Eastern blocks remains evident, as they consistently exhibit higher error values and larger mean logarithmic growth rates over time.

        \section{Conclusions}\label{sec:conclusions}
    This study investigates the predictability of Eastern and Western North Pacific atmospheric blocking events using an analogue-based approach. The results demonstrate that Eastern blocks exhibit lower predictability of their position compared to their Western counterparts. This is evident from the more rapid divergence of analogue-reference pairs in Eastern blocks, as measured by mean error evolution and the mean logarithmic growth rate, which quantifies sensitivity to initial perturbations \cite{lorenz1969atmospheric}. Due to limitations in the robustness of the analysis, no significant difference in persistence between eastern and western North Pacific blocks could be determined.
    
    The strength of this study lies in its sensitivity analysis, where different threshold values for blocking detection `$\alpha$' were tested from 0 to 1 to determine the most suitable values. The final selection of $\alpha= 0.20$ and $\alpha= 0.25$ ensures an optimal balance, where a sufficient number of blocking days are included while maintaining the synoptic-scale spatial coherence of blocking patterns. The study also prioritizes robust analogue selection by considering only the closest 50 and 75 reference-analogue pairs, ensuring that the selected analogues exhibit very close initial geopotential height distributions and locations. This approach thus provides a robust assessment of the intrinsic predictability differences between Eastern and Western blocks.

    The findings have implications for medium-range weather forecasting, particularly in regions influenced by North Pacific blocking, as the lower predictability of Eastern blocks contributes to increased forecast uncertainty. Additional studies of the impact of initial condition estimates on the predictability and skill of Eastern and western  blocking dynamics in Numerical weather prediction models (NWPs) are worth doing. 
    
    Despite the robustness of our results, certain limitations must be acknowledged. The study relies on a single climate model (MIROC6), which may introduce model-specific biases. Expanding the analysis to multiple CMIP6 models would enhance the generalizability of the findings. Another important way forward is to investigate the impact of climate change on the predictability of blocking using various projections.

    \backmatter
    \pagebreak
    \section*{acknowledgements}
    We would like to thank Francesco Ragone and Mireia Ginesta for the helpful discussions, and Jonathan Demaeyer for the technical support. Anupama K Xavier has received funding from the European Union’s Horizon 2020 research and innovation programme under the Marie Sklodowska–Curie grant agreement no. 956396.
    Oisín Hamilton has received funding from the European Union’s Horizon 2020 research and innovation programme under the Marie Sklodowska–Curie grant agreement no. 956170, as well as through the "Fédération Wallonie-Bruxelles" with the instrument "Fonds Spéciaux de Recherche".

    \section*{Competing Interests}
    The contact author has declared that none of the authors has any competing interests.

    \section*{Author Contribution}
    AKX contributed to conceptualisation, method development, method implementation and data analysis and writing and visualisation. OH contributed to conceptualisation, method development, writing and text improvements. DF and SV contributed to supervision, conceptualisation and writing

    \section*{Data availability}
    We used CMIP6 MIROC6 historical \cite{tatebe2018miroc} simulations for the study which can be downloaded from \href{https://aims2.llnl.gov/}{ESGF} metagrid website.

    \section*{Code availability}
    Code used for the study is in the process of being documented and will be released as open source before the conclusion of the review process.
    \bibliography{sn-bibliography}

\end{document}